# Nano-scale visualization of magnetic vortices in metal nanoparticles

*Satoko Toyama[1]*, Yoshiki O. Murakami[1], Ayako Nishikawa[1], Takehito Seki[1], Akihito Kumamoto[1,2], Yuichi Ikuhara[1,3,4], Naoya Shibata[1,3,5]**

[1]Institute of Engineering Innovation, School of Engineering, The University of Tokyo, Tokyo, Japan.

[2]JEOL Ltd., Akishima, Tokyo, Japan.

[3]Nanostructures Research Laboratory, Japan Fine Ceramics Center, Aichi, Japan.

[4]WPI Research Center, Advanced Institute for Materials Research, Tohoku University, Sendai, Japan.

[5]Quantum-Phase Electronics Center (QPEC), The University of Tokyo, Tokyo, Japan.

KEYWORDS: magnetic nanoparticle, magnetic vortex, scanning transmission electron microscopy, differential phase contrast, nano-scale imaging, magnetic field imaging

## Abstract

Magnetic vortices in individual nanoparticles are fundamental spin structures that govern the properties of next-generation magnetic devices and biomedical applications. However, directly imaging their complete structure, from the circulating in-plane magnetization to the nanometer scale out-of-plane core, remains challenging. Here, direct and quantitative visualization of magnetic vortex structures in individual cobalt nanoparticles is achieved by integrating a time-reversal methodology with tilt-scan-averaged differential phase contrast scanning transmission electron microscopy in a magnetic-field-free environment. This approach enables magnetic imaging in conjunction with atomic-scale analysis and reveals a correlation between particle geometry and internal demagnetizing fields. In addition, dynamic evolution of the vortex under in situ magnetic-field application is tracked, enabling unambiguous determination of the out-of-plane core polarity. This approach provides a powerful platform for correlating atomic-scale structure and magnetism within individual nanoparticles and for understanding the origins of nanoscale magnetic properties, thereby supporting the rational design of advanced nanomagnetic materials and devices.

The unique magnetic properties of individual nanoparticles are at the heart of numerous scientific and technological advancements, driving innovations from high-density data storage to targeted biomedical applications such as magnetic hyperthermia [1–4]. In nanoparticles of specific geometries, such as nanoplatelets, the delicate competition between exchange, magnetostatic and magnetic anisotropy energies can result in the formation of a stable ground state known as a magnetic vortex. This magnetic vortex consists of in-plane curling magnetization that surrounds a nanometer-scale, out-of-plane core. The collective behavior of this entire spin texture, from the circulating in-plane moments to the polarity of the central core, governs the response to external stimuli and defines its potential use in future information technologies [5–7].

However, characterizing this vortex state is often complicated by the fact that many experimental techniques measure the averaged response of a large ensemble of nanoparticles, a process that inevitably masks the distinct magnetic behavior of each particle. Therefore, direct, real-space visualization of the particle's magnetic vortex structure in conjunction with its atomic structure is essential for a complete understanding of its properties. The vortex configuration is exquisitely sensitive to subtle variations in a nanoparticle's specific size, shape, and crystallinity [8–10]. Experimentally obtaining both magnetic and crystal-structure information from an individual particle is therefore crucial for validating theoretical models that account for these factors and for the rational design of next-generation nanomagnetic applications.

A comprehensive understanding of the magnetic vortex in a nanoparticle requires satisfying several criteria simultaneously: (i) quantitative mapping of the magnetic field distribution with nanometer-scale resolution, (ii) simultaneous acquisition of the crystal structure and shape to correlate magnetic features with particle structure and morphology and (iii) the ability to probe the vortex in its native, zero-field state and to track its dynamic response to an applied external magnetic field. Despite progress with techniques such as scanning probe microscopy [11–14], X-ray microscopy [5,9,15], Lorentz microscopy [16], and electron holography [17–22], it has been difficult to fulfill all these demanding requirements for an individual particle. On the other hand, differential phase contrast scanning transmission electron microscopy (DPC STEM) has emerged as a promising approach [23,24], as it enables atomic-scale structural imaging and high-resolution magnetic field mapping within an individual particle. However, its successful application to magnetic nanoparticles has been impeded by two obstacles. First, conventional high-resolution S/TEM requires placing the specimen within the strong magnetic field of the objective lens, which can perturb or even destroy the native vortex state. Second, the phase shift of the electron beam caused by the weak magnetic field is fundamentally superimposed on a signal from the specimen's electric field, which is dominated by the mean inner potential (MIP) [25] in nanoparticles. A time-reversal method offers a way to separate the electric and magnetic field signals [20,26,27]. The technique relies on physically flipping the specimen to reverse the electron beam's direction relative to the sample. This reversal preserves the direction of electric field deflection but inverts that of the magnetic field deflection, allowing the two contributions to be separated in principle by subtraction of one from the other. However, its practical implementation in DPC STEM has been unreliable. This is because flipping the sample introduces slight but unavoidable changes in the crystal's orientation, which generate non-negligible dynamical diffraction artifacts, hindering the observation of the true magnetic signal.

Here, we overcome these long-standing challenges by implementing a time-reversal methodology within a tilt-scan-averaged DPC (tDPC) [28–30], all integrated into a magnetic-field-free-atomic-resolution STEM [31]. The tDPC technique is particularly powerful as it averages out the dynamical diffraction artifacts that plagued previous attempts; indeed, it has proven successful in quantitatively observing electromagnetic fields near crystalline defects such as grain boundaries [32], heterointerfaces[33,34], and magnetic and polar domain walls [35,36]. This synergistic approach allows us to robustly separate the magnetic field information from the electric field background, as illustrated in **Figure 1**. In the present study, we demonstrate direct and quantitative visualization of the complete magnetic vortex structure in individual cobalt nanoparticles and track the vortex core response to an externally applied magnetic field through in situ observation.

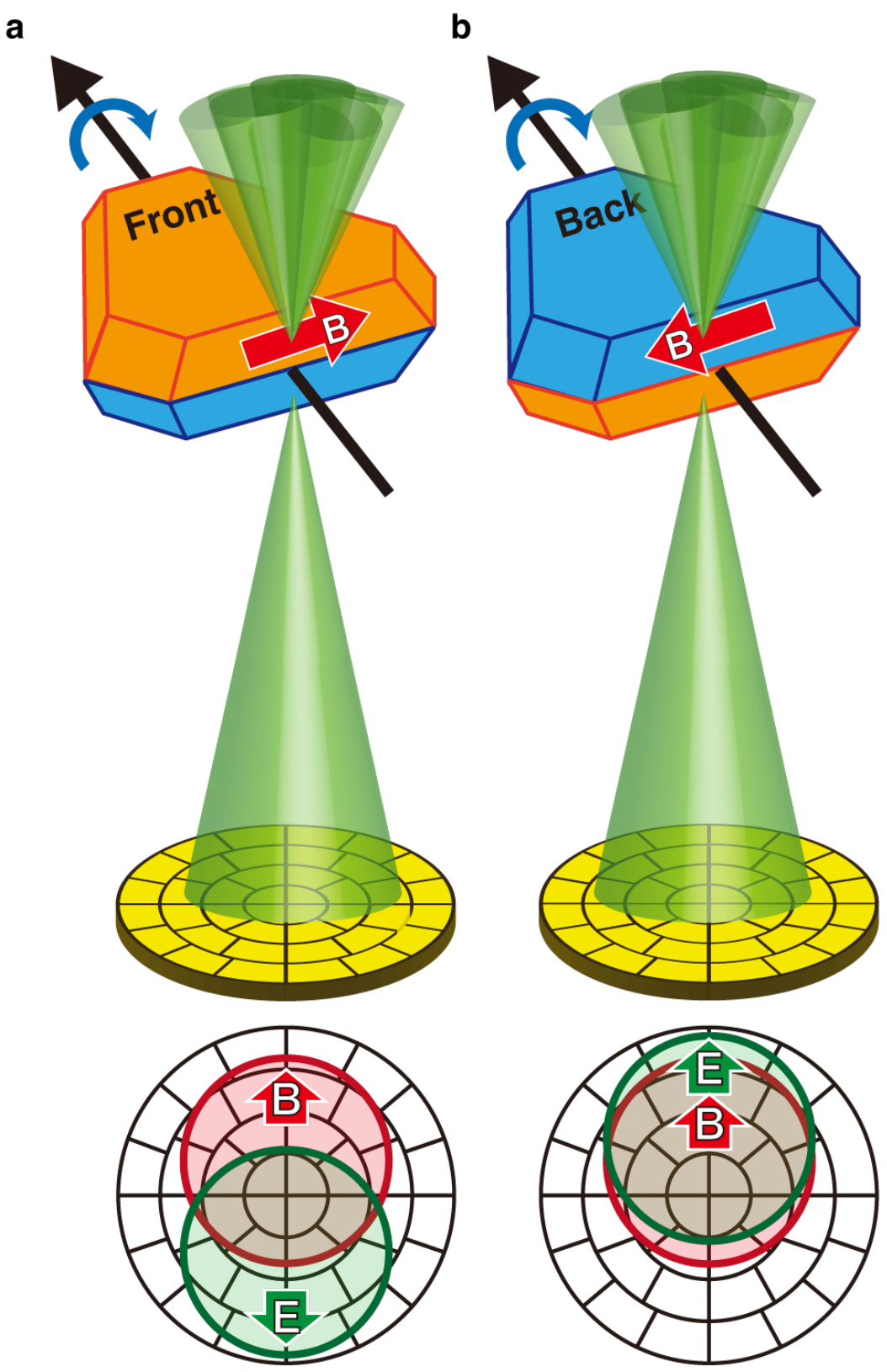

**Figure 1.** Schematic of the time-reversal tDPC STEM method for separating electric and magnetic fields. (a) In the "Front" observation, the beam deflection of the transmitted electron beam on the detector is a superposition of contributions from the electric field, E (green disk) and the magnetic field, B (red disk). (b) After flipping the specimen for the ‘back’ observation, the electron beam’s path relative to the sample is effectively reversed.

The cobalt nanoparticles were synthesized via a gas evaporation method [37] on a thin amorphous silicon support film. High-purity cobalt was then evaporated from a Joule-heated tungsten boat within an Ar-$H_2$ gas atmosphere, allowing the vaporized atoms to instantaneously condense and undergo coalescence growth, resulting in the formation of soot-like crystalline nanoparticles that were transported by thermal convection and deposited onto the upward-facing support film.

STEM observations, including HAADF and tDPC experiments, were conducted using a magnetic-field-free atomic-resolution STEM at The University of Tokyo (JEM-ARM200F-based MARS, JEOL Ltd.). For atomic-scale HAADF-STEM observation, the accelerating voltage and convergence semi-angle were set to 200 kV and 20 mrad, respectively. The final high-S/N image was obtained by averaging 30 images using normalized cross-correlation.

**Figure 2** shows the structural characterization of the obtained nanoparticles. HAADF STEM images show that the nanoparticles have triangular-like (Figure 2a) and hexagonal-like (Figure 2b) in-plane morphologies. Although bulk cobalt typically has an hcp structure at room temperature, the high-temperature fcc phase is stabilized even at room temperature in these nanoparticles due to their reduced size [38,39], as confirmed by an atomic-scale HAADF image (Figure 2d). The nanoparticles are truncated plates exposing a large (111) surface, as depicted in the structural model (Figure 2c), and contain a twin boundary at the middle of their thickness. Based on the geometry of the truncated regions and position-averaged convergent beam electron diffraction analysis, the thicknesses of the triangular-like and hexagonal-like particles were estimated to be approximately 50 nm and 45 nm, respectively. These structures are consistent with earlier reports [38,40]. Individual nanoparticles with these triangular and hexagonal shapes were selected for subsequent magnetic field imaging.

We then performed tDPC STEM observation to extract magnetic structures in these particles. In tDPC STEM, the in-plane electromagnetic field inside a specimen imparts a momentum transfer to the transmitted electron beam, resulting in a deflection angle vector, $\Theta$. Under the phase object approximation, this deflection is proportional to the path-integrated Coulomb and Lorentz forces:

$$\Theta(x, y) = -\frac{e\lambda}{hv}\int(\boldsymbol{E}_{\perp} + \boldsymbol{v} \times \boldsymbol{B}_{\perp})\mathrm{d}z \quad (1)$$

where e is the elementary charge, $\lambda$ is the electron wavelength, h is the Planck constant, $\boldsymbol{v}$ is the electron velocity vector, and $\boldsymbol{E}_{\perp}$ and $\boldsymbol{B}_{\perp}$ are the in-plane components of the electric and magnetic fields, respectively, integrated along the beam path z. For the cobalt nanoparticles observed in this study, which are uncharged, the electric field $\boldsymbol{E}_{\perp}$ is assumed to arise solely from the gradient of the MIP.

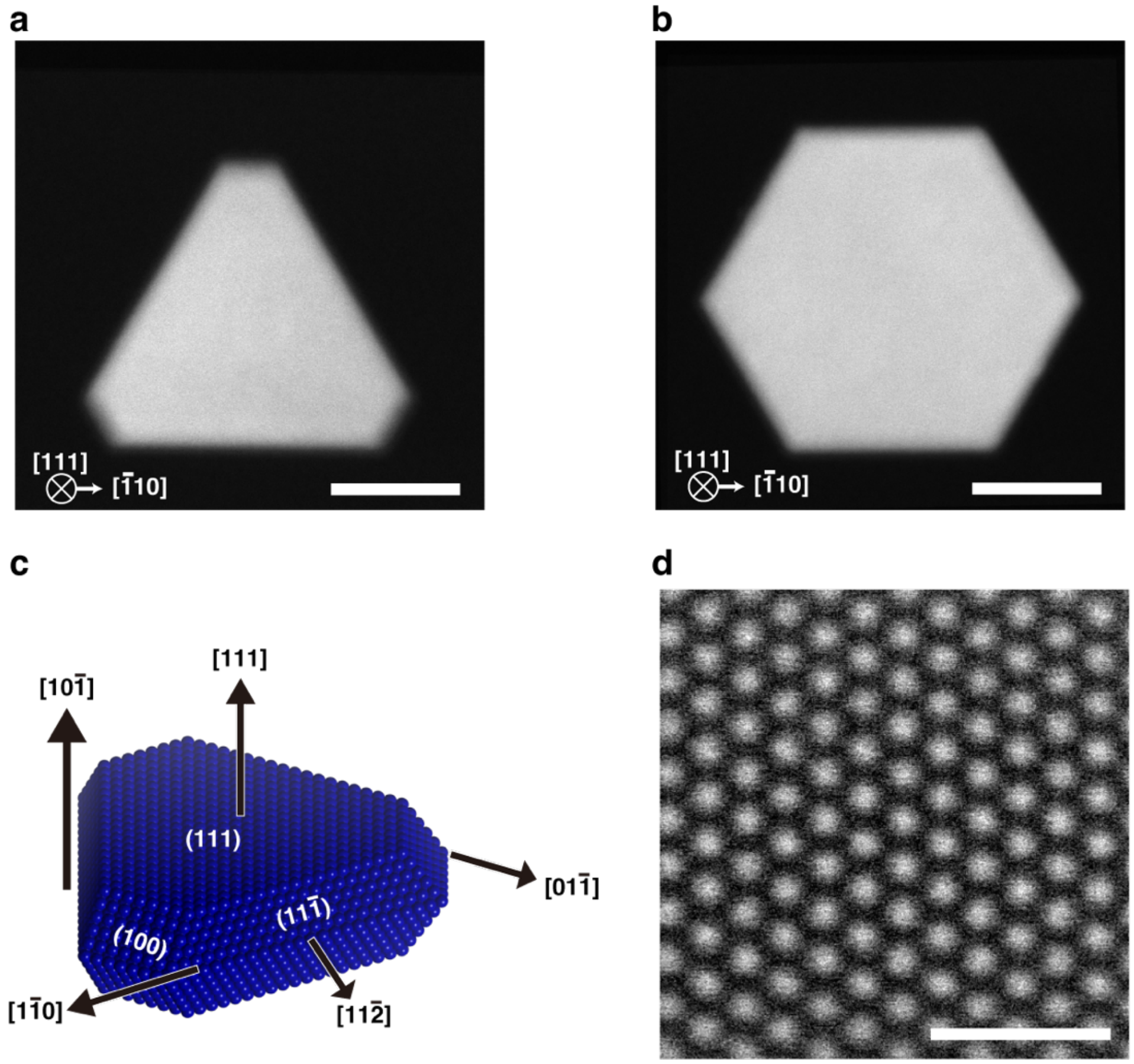

**Figure 2.** Structural characterization of cobalt nanoparticles. (a)(b) High-angle annular dark-field (HAADF) STEM images showing nanoparticles with triangular (a) and hexagonal (b) morphologies. (c) Atomic model of a truncated fcc nanoparticle, illustrating the large (111) surface and crystallographic orientations. (d) Atomic-resolution HAADF-STEM image confirming the fcc crystal structure. Scale bars, 50 nm (a, b); 1 nm (d).

The time-reversal tDPC STEM method separates electric and magnetic fields by their different symmetries under reversal of the electron beam velocity. Flipping the specimen by 180° inverts the magnetic deflection while leaving the electric deflection unchanged. This "back" observation is equivalent to reversing the velocity vector ($\boldsymbol{v} \rightarrow -\boldsymbol{v}$) relative to the specimen's internal fields. The deflection equation for the back observation, $\Theta_{\mathrm{back}}$, then becomes:

$$\Theta_{\mathrm{back}}(x, y) = -\frac{e\lambda}{hv}\int(\boldsymbol{E}_{\perp} - \boldsymbol{v} \times \boldsymbol{B}_{\perp})\mathrm{d}z \tag{2}$$

Note that the electric field term remains unchanged, while the magnetic field term, arising from the cross product with $\boldsymbol{v}$, inverts its sign. Therefore, by computationally adding and subtracting the deflection maps from the front ($\Theta_{\mathrm{front}} \equiv \Theta$) and back observations, the pure electric and magnetic signals can be isolated:

$$\Theta_{\mathrm{front}} + \Theta_{\mathrm{back}} \propto \int \boldsymbol{E}_{\perp}\mathrm{d}z \tag{3}$$

$$\Theta_{\mathrm{front}} - \Theta_{\mathrm{back}} \propto \int \boldsymbol{B}_{\perp}\mathrm{d}z \tag{4}$$

The tDPC STEM observations were performed with an accelerating voltage of 200 kV and a convergence semi-angle of 1 mrad, which corresponds to a probe size of approximately 1 nm. Tilt-scan averaging was performed using 61 beam tilts with a maximum tilt angle of 3.5 mrad, under the detector condition shown in Fig. S1. To enhance the S/N, five images of the same field of view were acquired and subsequently averaged using a cross-correlation algorithm. For each acquired image, a beam deflection map was obtained by calculating the approximated center of mass (CoM) of the diffraction pattern using signals from the segmented detector [41]. A critical step in the time-reversal method is the precise alignment of the 'front' and 'back' tDPC images, as scan distortions can cause relative image shifts and distortions. To correct for this, we performed image registration using an affine transformation. Three reference regions were selected from the supporting amorphous Si film surrounding the nanoparticle (selected area is shown in Figure S2). An affine transformation matrix for the 'back' image was then optimized using a least-squares method to maximize the normalized cross-correlation between the reference regions of the 'front' and 'back' images. After registration, the pure electric and magnetic deflection maps were extracted by adding and subtracting the aligned 'back' image from the ‘front’ image. A discrete cosine transformation was then performed to generate a magnetic phase image, which also served to denoise the magnetic field images [42].

**Figure 3**a-h shows the front and back beam deflection images acquired from the triangular-like and hexagonal-like particles. We define the 'front' orientation as beam incidence along the $[111]$ direction, and 'back' as the reverse orientation along the $[\bar{1}\bar{1}\bar{1}]$ direction after flipping the specimen. The resulting electric and magnetic deflection maps are shown in Figure 3i-p. A strong edge contrast, identical in both front and back images, appears exclusively in the electric-field maps (Figure 3m-p). This contrast is attributed mainly to variations in the mean inner potential (MIP) at the tapered edges of the nanoparticles. A nanometer-scale granular contrast, also visible in both front and back images, is present exclusively in the electric-field map similarly to the edge contrast. This feature indicates that the granular contrast is an electrostatic field component, resulting from variations in the density and/or thickness of the supporting amorphous Si film. By contrast, the black and white contrast spreading across the entire particles is reversed between the front and back images (Figure 3a-h) and extracted solely in the magnetic-field maps (Figure 3i-l), which represent the circulating magnetic fields, namely magnetic vortices.

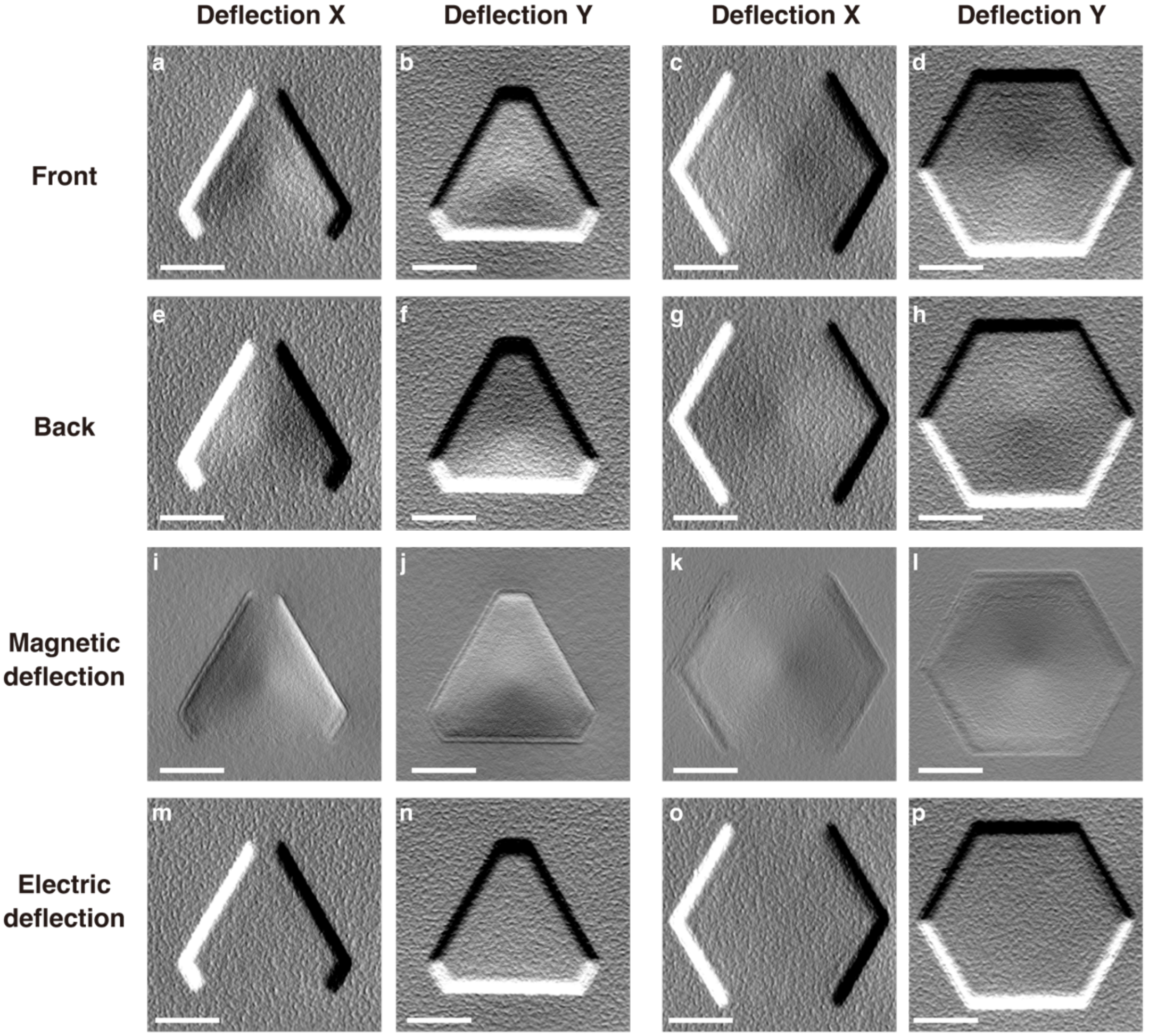

**Figure 3.** Separation of electric and magnetic deflection maps using the time-reversal tDPC method. (a-h), Raw beam deflection maps in the X and Y directions for the triangular and hexagonal particles, acquired from the 'front' (a-d) and 'back' (e-h) orientations. (i-l), Isolated magnetic deflection maps obtained by subtracting the 'back' images from the 'front' images. A clear vortex pattern is now visible within each nanoparticle. (m-p), Isolated electric deflection maps obtained by adding the 'front' and 'back' images. All scale bars correspond to 50 nm.

**Figure 4a-f** show the processed images: (a,d) magnetic field vector maps, (b,e) field-strength images, and (c,f) phase images for the two particles. In both cases, a clear vortex circulation and a distinct core region are visualized. The magnetic phase images (Figure 4c, f) clearly resolve the differences in the phase shift arising from the distinct particle geometries; the phase within the triangular particle demonstrates a distribution that is more closely aligned with a triangular configuration whereas the phase within the hexagonal particle exhibits an almost circular distribution. Furthermore, a notable difference appears in the field-strength maps of the two particle shapes. Specifically, the triangular particle (Figure 4b) shows three distinct lines of suppressed magnetic field strength radiating from the central core toward each vertex. This feature is not as evident in the more symmetric hexagonal particle (Figure 4e).

To understand this feature, we performed micromagnetic simulations using the MuMax3 software[43]. For this simulation, the three-dimensional shape of the nanoparticles was modeled based on the HAADF STEM images, assuming a truncated particle structure. A mesh size of 1 $nm^3$ was used for all simulations. The material parameters for cobalt were adopted from the previous studies [44,45]. The simulation results, shown in **Figure S3**, reproduce the experimental observations and reveal that these linear suppressions distinct in the triangular-like particle are a direct consequence of the demagnetizing field. Magnetostatic energy in the nanoparticles is minimized by forming flux-closure magnetization patterns that prevent effective magnetic charges from appearing at their edges. At the acute vertices in the triangular particle, the magnetization needs to bend sharply, generating effective magnetic charges associated with the local demagnetizing field. This field thereby generates the pronounced linear suppression features in the triangular particle. The different vertex angles of the triangular and hexagonal particles therefore determine the demagnetizing-field strength in each case.

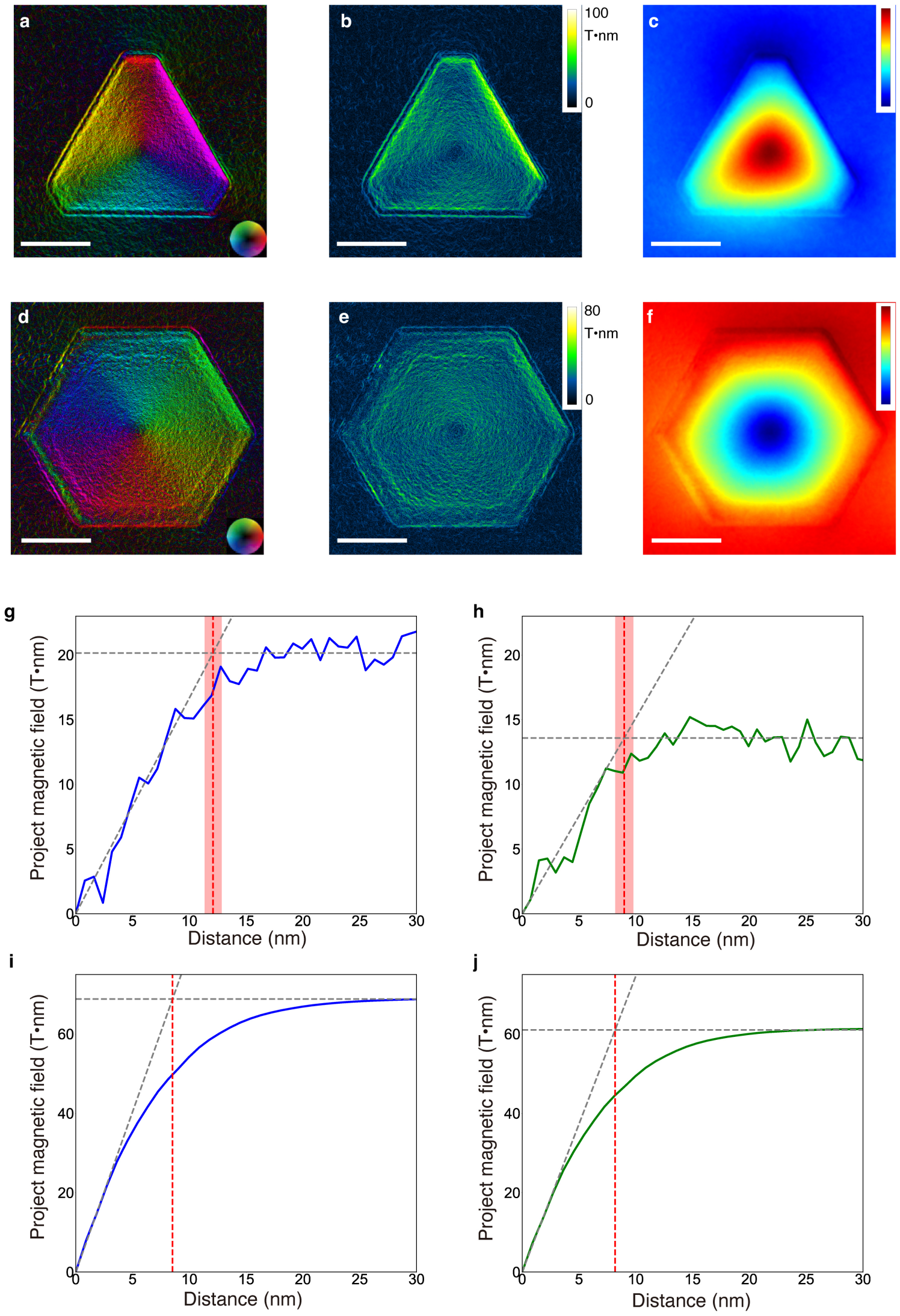

**Figure 4.** Figure 4. Reconstructed magnetic fields and vortex core size analysis of cobalt nanoparticles. (a–c) Magnetic field vector map, field-strength map, and magnetic phase shift image for the triangular-like particle, showing a clockwise vortex. (d–f) Corresponding maps for the hexagonal-like particle, showing a counter-clockwise vortex. (g–j) Experimental and simulated radial profiles of the projected magnetic field strength for the triangular-like (g, i) and hexagonal-like (h, j) particles. The core radius is defined by the maximum-slope position, indicated by the red dashed line. Shaded regions in (g) and (h) indicate the statistical error of the linear fitting. Scale bars, 50 nm.

For the analysis of the magnetic core size, the field strength images were first smoothed using a Gaussian filter with a full-width at half-maximum (FWHM) of 0.5 nm. The vortex core profiles were then obtained by plotting the radial profile of the smoothed field-strength maps as a function of distance from the core's center (**Figure 4**g-j). As seen in the field-strength maps (Figure 4b and Figure S3b), the demagnetizing field in the triangular particle extends into the core region, causing its radial profile to appear broader than that of the hexagonal particle. As a result, the measured core diameter was 12.1±0.7 nm for the triangular particle and 9.0±0.8 nm for the hexagonal one. These features arising from the particle geometries may also affect the core stability. Indeed, our simulations show that the external field required to reverse the core magnetization is smaller for the triangular-like particle than for the hexagonal-like one (**Figure S4**), suggesting that the triangular-like particle's vortex is less stable and easier to switch. Furthermore, the experimental core sizes are consistently larger than those obtained from the simulations. This discrepancy may be attributed to surface oxidation of the particles, which can reduce the projected magnetic signal and alter the magnetic properties from those of pure cobalt.

Finally, we performed in-situ tDPC observation under external magnetic fields to determine the out-of-plane vortex core polarity. A key advantage of using magnetic-field-free STEM is that the particles can be observed in their pristine, as-prepared magnetic state. In this zero-field condition, the 'up' and 'down' core polarities are energetically equivalent, meaning that either state is equally possible. While our time-reversal experiments reveal the static in-plane magnetic vortex circulation, a key challenge in nanomagnetic materials is to track the dynamic evolution of such vortex core structures under an external magnetic field. We demonstrate this crucial capability by resolving the out-of-plane core polarity, which is a parameter that governs the vortex's dynamic response but cannot be determined by static observation. To achieve this, we performed in-situ tDPC experiments on the triangular-like particle by applying out-of-plane magnetic fields of ±200 mT. The positive field direction is defined as being parallel to the $[\bar{1}\bar{1}\bar{1}]$ direction of the particle. As illustrated in **Figure 5**a, the application of an external magnetic field provides a direct means of assigning the core polarity. When an external field is applied parallel to the core magnetization, more spins tilt out of plane, expanding the core size, which can lead to a suppressed radial profile of the in-plane magnetic field. Conversely, an antiparallel field shrinks the core size, forcing more spins into the in-plane and resulting in an enhanced in-plane profile.

Because the MIP-induced electric field is expected to remain unchanged under the applied magnetic field, we isolated the magnetic contribution in each in-situ image by subtracting the electric-field map measured at zero magnetic field (Figure 3m-p). The experimental radial profiles of the projected magnetic field strength (**Figure 5**b) clearly show the field-induced response of the vortex core: the +200 mT field suppresses the in-plane profile, while the -200 mT field enhances it. This dynamic response unequivocally identifies the core polarity as pointing parallel to the +200 mT field direction, in agreement with micromagnetic simulations (**Figure 5**c). This successful determination of the core polarity serves as a clear validation of our method's ability to directly visualize the subtle, field-induced evolution of magnetic structures at the nanoscale.

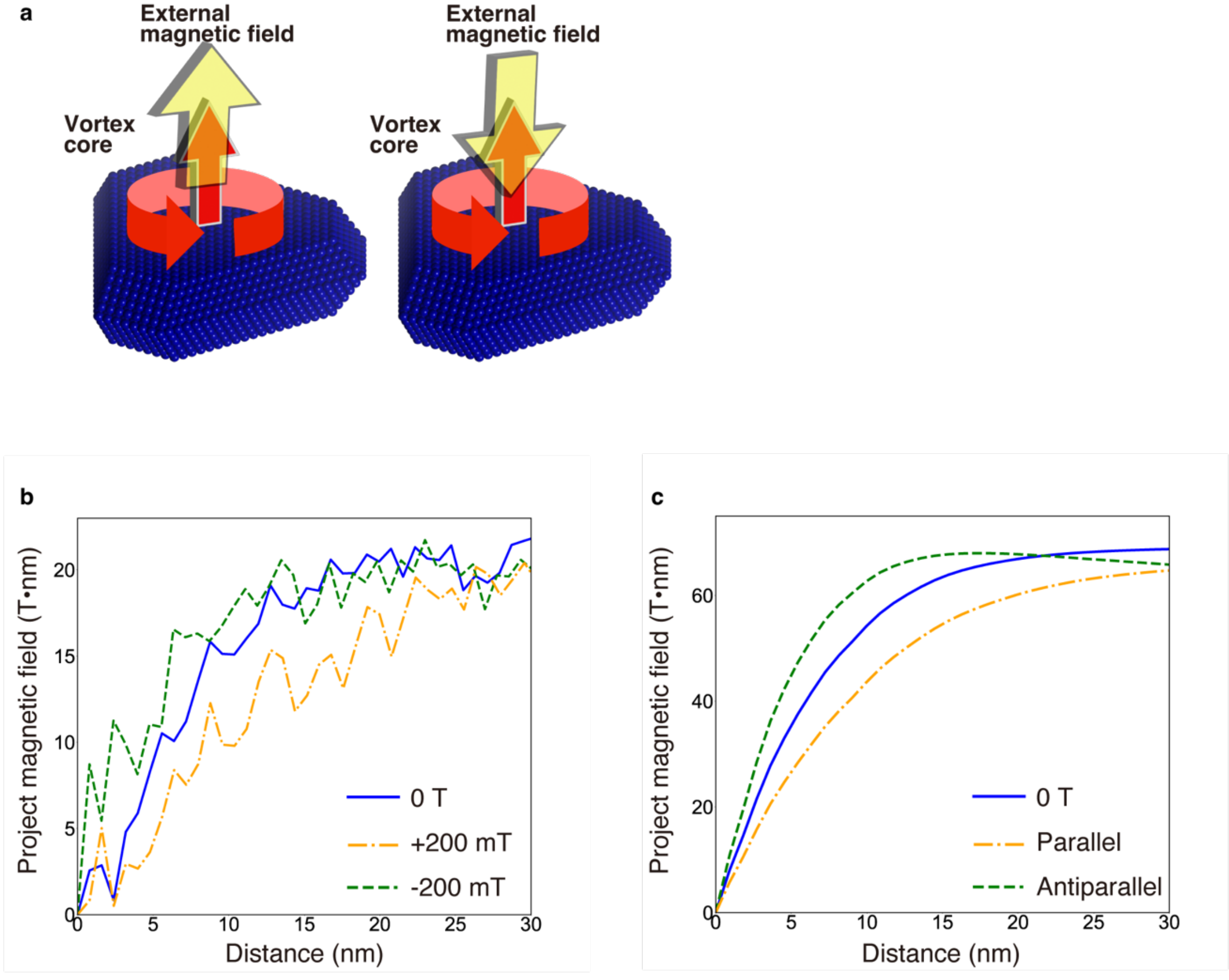

**Figure 5.** In-situ observation of vortex core under external magnetic field. (a) Schematic illustrating the effect of an external out-of-plane magnetic field. A field applied parallel to the core magnetization (left) is expected to expand the core, while an antiparallel field (right) should shrink it. (b) Experimental radial profiles of the projected in-plane magnetic field strength for the triangular nanoparticle under zero field (0 T), +200 mT, and -200 mT. (c) Corresponding profiles obtained from micromagnetic simulations.

In summary, we have demonstrated the quantitative, real-space visualization of magnetic vortex structures in individual nanoparticles. By developing and implementing a time-reversal methodology in tDPC STEM under magnetic-field-free conditions, we have overcome the challenges of separating magnetic signals from electric-field backgrounds in individual nanoparticles while suppressing dynamical diffraction artifacts. This approach provided not only a static magnetic field map of the in-plane magnetic vortex circulation and its correlation with nanoparticle geometries, but also enabled the direct visualization of its dynamic evolution under an external magnetic field, leading to the successful determination of the out-of-plane core polarity. These capabilities establish time-reversal tDPC STEM as a direct probe of structure–magnetism correlations and field-induced vortex dynamics in individual nanomagnets.